\documentclass[12pt,epsfig,epsf]{article}
\usepackage{epsfig}
\begin{document}
\def\cst{CS$_2$}
\title{GEM Operation in Negative Ion Drift Gas Mixtures}
\author{J. Miyamoto, I. Shipsey, Purdue University\\
 C. J. Martoff, 
M. Katz-Hyman, R. Ayad, Temple University\\
 G. Bonvincini, A. Schreiner,
Wayne State University}
\maketitle
\abstract{
The first operation of GEM gas gain elements in negative ion gas mixtures 
is reported.  Gains up to several thousand were obtained from single-stage
GEMs in \cst\, vapor at low pressure, and in mixtures of  \cst \,with Argon and 
Helium, some near 1 bar total pressure.}

\section{Introduction}
Microstructured detectors such as GEMs \cite{cern,shipsey} and Micromegas
\cite{cern2,shipsey2} are seeing rapidly increasing application to 
physics experiments and applied problems\cite{somereview}.  These detectors
offer a number of advantages over other gas gain structures, including
ruggedness, mass manufacture-ability, high speed, and
greatly reduced positive ion backstreaming in TPC applications.  

Negative ion drift gases (NI-gases) are another innovation in gas detector
technology which are just beginning to become known\cite{negion,uclc,drift,TPCSymp}.
Certain electronegative gases allow the primary ionization in a drift or TPC
device to be transported to the gain elements in the form of negative
molecular ions\cite{Crane}.  Mainly due to the mass matching of the
drifting ions with 
respect to the gas molecules, the ions are much more tightly thermally coupled
to the gas than are drifting electrons.  As a result,
the drift-diffusion, {\it both longitudinal and transverse}, remains at
the thermal (lower) limit up to extraordinarily high drift fields, several
tens of V/cm$\cdot$Torr at least.  This can give a dramatic advantage in 
space-point resolution, particularly in long drift geometries and/or
where imposing the usual magnetic field along the drift direction would
be impractical or undesirable.  Of course, ions of mass $m_I$ will 
drift approximately $\sqrt{m_I/m_e} \sim $ 500 times slower than 
electrons at the same 
reduced field.  This is usually a disadvantage but in low rate experiments
with high channel count, it may actually be a significant advantage.  
The spatial
resolution (particularly in the drift direction) can still be quite 
high, even using very 
low-bandwidth and hence low-noise electronics.

The present work demonstrates the compatibility of GEM gas amplification 
with negative ion drift. Furthermore it describes for the first time 
some results for GEM gain in NI-gas mixtures near 1 bar total pressure.
Such mixtures will considerably ease the construction of large NI-gas
detectors by removing the need for operation at reduced pressure in a 
vacuum vessel.  Drift velocities and longitudinal diffusion have
also been 
measured for these NI-gas mixtures, and will be  reported on separately.
However the measured drift velocities in all the mixtures reported on here,
do show pure NI-gas behavior under the conditions of these measurements.

\section{Apparatus}
The apparatus used in the present work is shown schematically in Figure 1. 
A single 50 mm diameter GEM manufactured by 3M Corporation\cite{3mgem} was 
mounted 7.5 mm below a transparent (mesh) electrode, within a stainless steel 
bell jar vacuum system used as a gas envelope.  Drift voltages of 
up to -500 Volts
were applied to the mesh in the measurements reported below.  The top electrode
of the GEM was operated at ground potential, and the bottom electrode at 
variable positive voltages ranging up to 580 V.  The GEM amplification signal
was read directly from the bottom electrode through an Ortec 142PC preamp
and an Ortec 572 shaping amplifier.  Shaper gains from 20-200 and
shaping times from 6 to 10 $\mu$s were used. The detector was irradiated by a collimated 
$^{55}$Fe x-ray source a few cm above the mesh.  The x-rays were 
directed downward along the drift direction near 
the center of the GEM area.  The source was opened and closed remotely.
Pulse height spectra were
obtained using an ORTEC ADCAM Analyst MCA.

\begin{figure}[h]
\begin{center}\thicklines
\epsfig{file=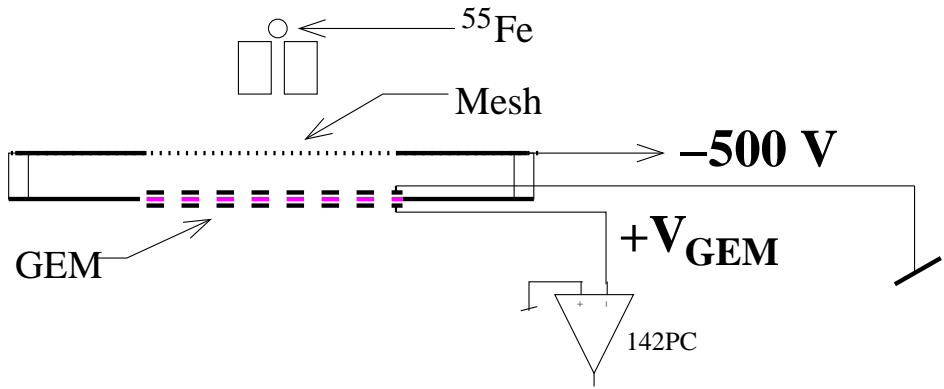,height=2.0in}
\caption[Schematic of Apparatus]{GEM apparatus including drift mesh grid,
GEM foil, collimated $^{55}$Fe source, and bias voltages.}
\end{center}
\end{figure}

To operate, the bell jar was closed and evacuated, backfilled with the desired
gas mixture (minor component first), and data were taken with the system sealed.
The rate of pressure rise was less than 50 mTorr/hour at the base pressure of 80
mTorr.  It should be noted that with negative ion gases the influence of 
air contamination is minimal, since capture by oxygen is a comparatively
weak, three body process\cite{o2cap}, and negative molecular ions of 
water are promptly stripped by drift fields of just a few tens of
V/cm\cite{waterdiss}.

\section{Results}
The results are summarized in Table 1 and Figure 2.  Several gas mixtures 
were tested, A single measurement was also made with Argon-Isobutane 
(a conventional ``e-gas").  These
are shown in Table 1 along with the gas gain achieved when each measurement 
was terminated, the logarithmic
slope of the gain vs. V$_{GEM}$ curve, and the last (highest) value of 
V$_{GEM}$ which was
tried.  Each measurement was terminated somewhat arbitrarily at a
V$_{GEM}$ that gave a satisfactorily high gain.
While no sparking or breakdown was observed in
the Ni-gas mixtures at any applied voltage, no attempt was made to find
the maximum sustainable value of V$_{GEM}$ for the different gases.

Pulse height spectra for two of the NI-gas mixtures are shown in Figure 3.
Note the enhanced Sulfur fluorescence-escape peak in the Helium mixture due
to the low x-ray attenuation of this mixture compared to the dimensions
of the drift gap.  The pulse height resolution in the argon mixture is 
about 35\% FWHM.  Electronic noise sets in at different levels in
the two spectra mainly because of the different shaping amplifier
gain settings used.

\begin{figure}[h]
\begin{center}\thicklines
\epsfig{file=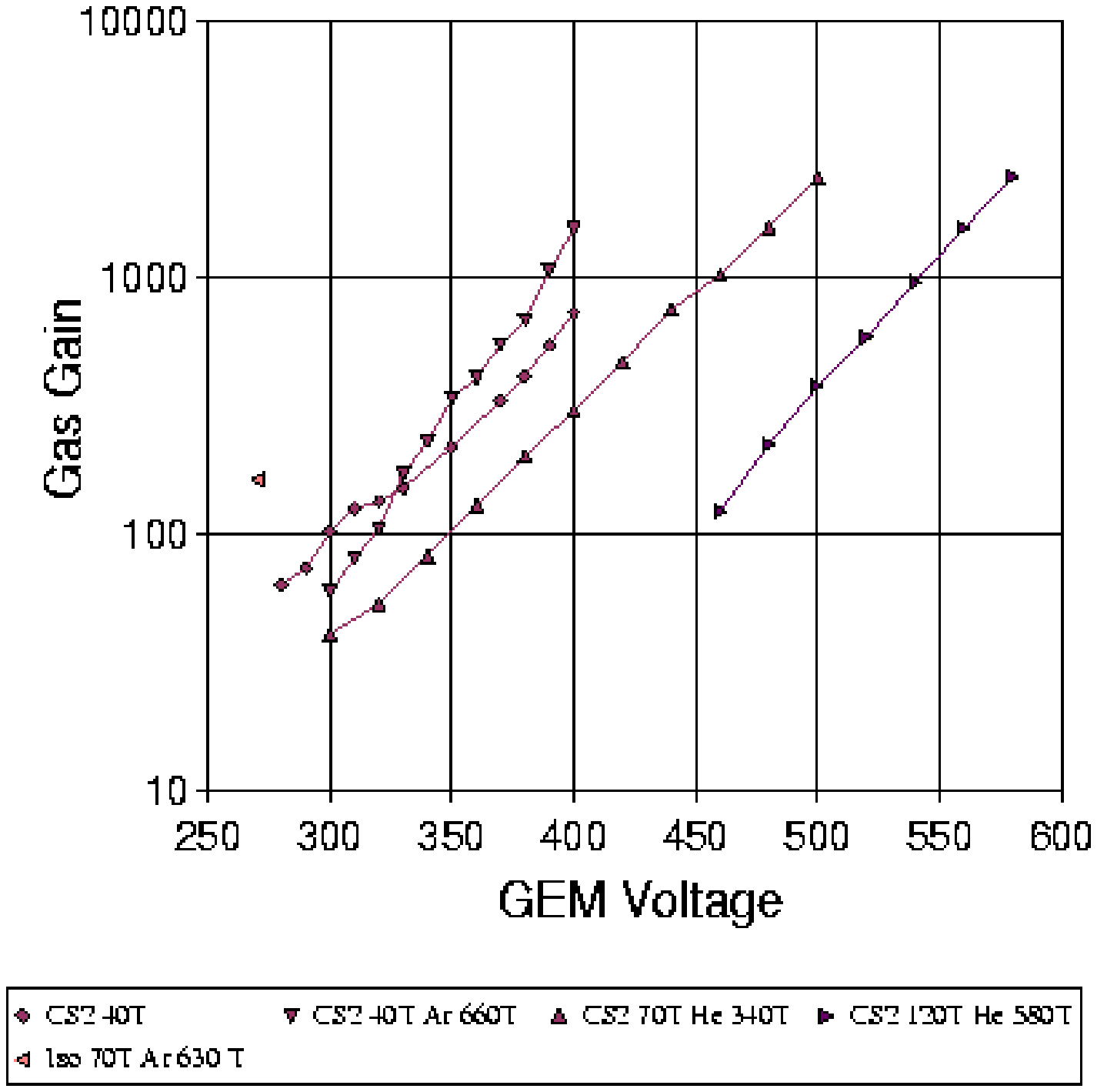,width=5.0in}
\caption[3M-GEM Gain Curves]{Gain vs. V$_{GEM}$ curves for
3M-GEMs exposed to collimated $^{55}$Fe source. The single point
at 270 V is for the e-gas Ar-Isobutane.} 
\end{center}
\end{figure}

\begin{table}[h]
\begin{tabular}{|l|l|l|l|}
\hline
Gas Mixture & V$_{GEM, max}$ & Max Gas Gain & $k$ \\
\hline
\hline
Ar 70 T+Iso 630T & 270 V & 162 & - \\
\hline
\cst\, 40 T & 400 & 729 & .019 \\
\hline
\cst\, 40 T+Ar 660T & 400 & 1540 & .032 \\
\hline
\cst\, 70 T+He 340 T & 500 & 2450 & .021 \\
\hline
\cst\, 120 T+He 580 T & 580 & 2460 & .025 \\
\hline
\end{tabular}
\caption[GEM gas gains]{Gas gains obtained with 3M-GEM in various
gas mixtures.  Only one point (270 V) was taken for the e-gas Ar/Iso.
The curves were terminated once a satisfactory gain was achieved;
no attempt was made to determine the maximum voltage the GEM would take.
No instability or sparking was observed at any voltage with any gas
mixture.  The last column gives the fitted logarithmic slope of the
gas gain vs. GEM voltage curves (Gain = $A \exp{k V_{GEM}}$).}
\end{table}

\begin{figure}[h]
\begin{center}\thicklines
\epsfig{file=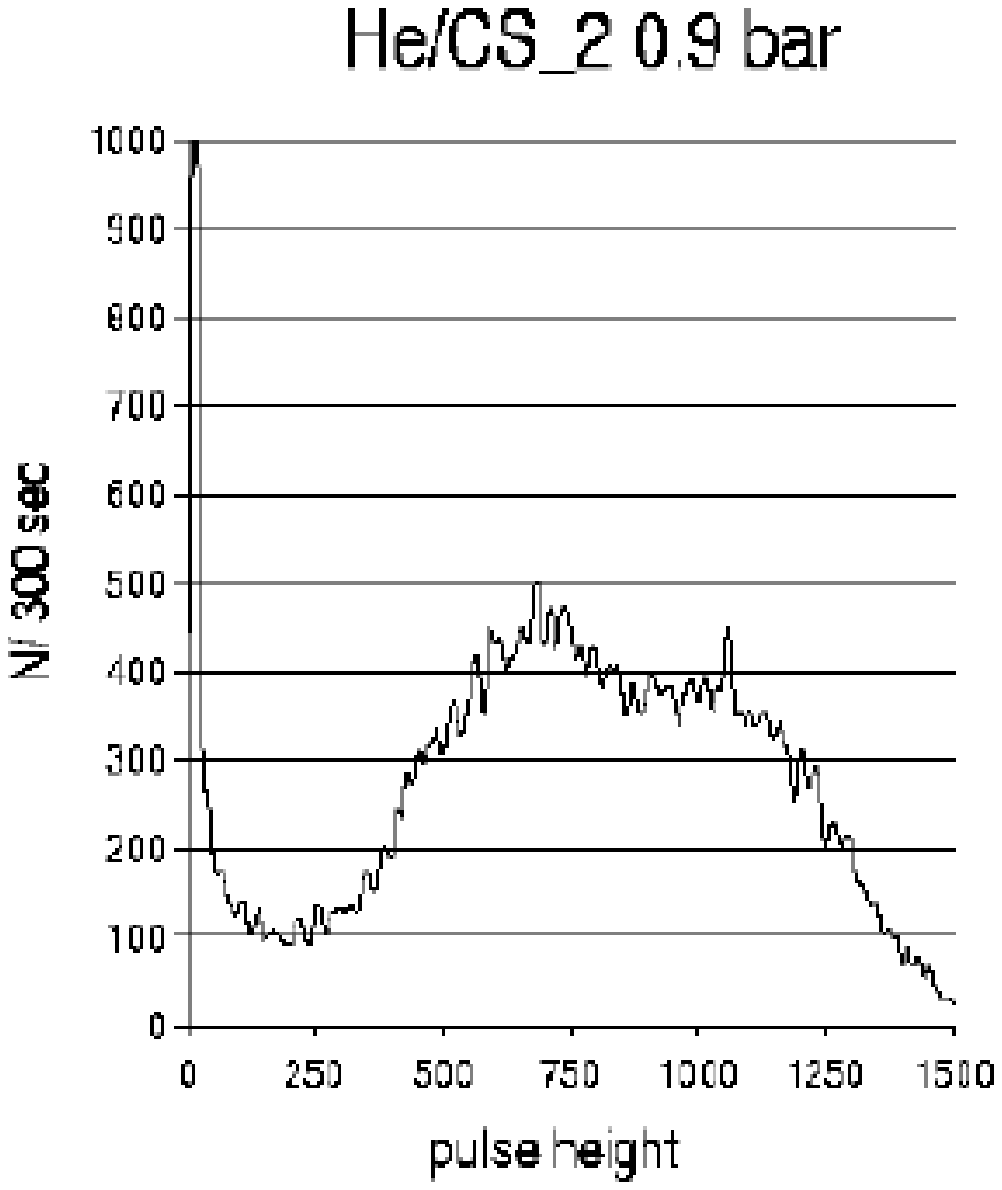,width=3in}
\epsfig{file=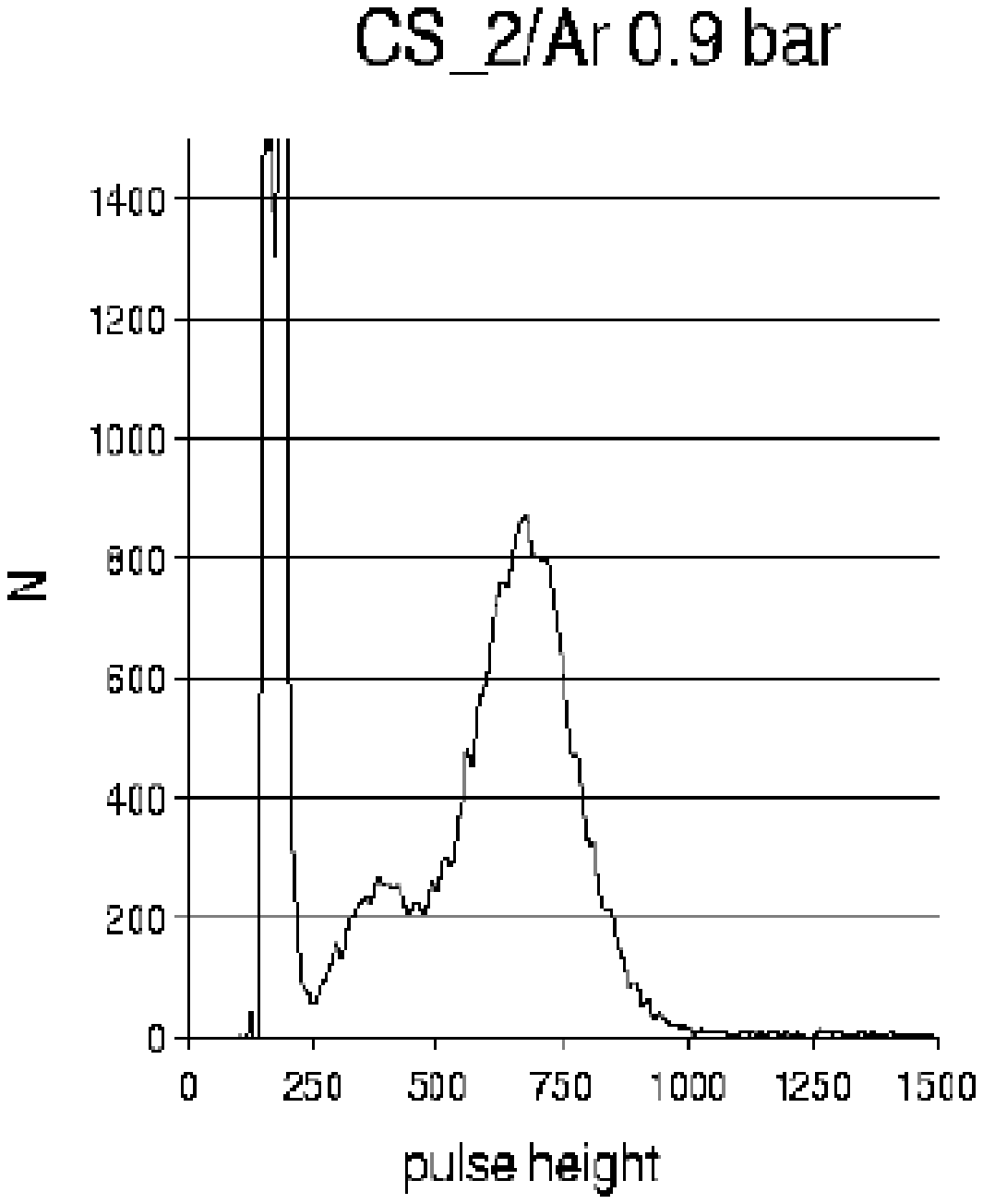,width=3in}

\caption[3M-GEM NI-Gas Pulse Ht. Spectra]{Pulse height spectra for
3M-GEMs exposed to collimated $^{55}$Fe source. Upper panel shows 
0.9 bar Helium mixture from Table 1, V$_{GEM}$=500 V, V$_D$= 375 V, 
shaper gain 20. 
Lower panel shows 0.9 bar Argon-Isobutane mixture from Table 1, 
V$_{GEM}$=375 V, V$_D$=500 V, shaper gain 200.}
\end{center}
\end{figure}

\section{Discussion}
GEM's in NI-gas mixtures show stable operation at moderately high gain.
GEM voltages up to 580 Volts were explored without any evidence of
sparking or instability.  

The near 1-bar Helium mixtures are of particular interest for gas-based
direction-sensitive Dark Matter searches.  Note that the
Helium component of the 1-bar 
\cst-Helium mixture only increases the total electron density
by 25\% over that of the \cst\, component alone. Thus a detector could
be constructed with low total gas density (hence reasonably long tracks
from low energy recoils, for direction determination), but operating
at or near 1 bar total pressure. This would permit
such experiments to be operated without a vacuum vessel, with its attendant
expense and radioactivity.  Other advantages of such a scheme would include
the excellent spatial resolution in all three dimensions afforded by 
negative ion drift, 
greatly reduced sensitivity to electronegative contaminants,
and a significant and variable content of medium-mass nuclei to 
kinematically match the favored range of WIMP masses.

\end{document}